\documentclass{aa}

\usepackage{graphicx}
\usepackage{txfonts}
\usepackage{amsmath,amssymb}
\usepackage{mathrsfs}   % include it to use \mathscr font
\usepackage{mathtools}  % for subscripted mapsto
\usepackage{color}
\usepackage[bookmarks=true]{hyperref}
\hypersetup{
  colorlinks=true,
  pdfborder=2 2 0.,
  linkcolor=red,
  anchorcolor=black,
  citecolor=blue,
  urlcolor=black,
}
\usepackage{breakurl}

\usepackage{natbib}
\bibpunct{(}{)}{;}{a}{}{,} % to follow the A&A style

\newcommand{\Msun}{M_\odot}
\newcommand{\de}{{\rm d}}
\newcommand{\bx}{{\bf x}}
\newcommand{\bv}{{\bf v}}
\newcommand{\bJ}{{\bf J}}

\newcommand{\bw}{{\bf w}}

\newcommand{\Jr}{J_R}
\newcommand{\Jphi}{J_\phi}
\newcommand{\Jz}{J_z}

\newcommand{\sigmar}{\sigma_R}
\newcommand{\sigmaz}{\sigma_z}
\newcommand{\sigmaphi}{\sigma_\phi}
\newcommand{\Rc}{R_{\rm c}}
\newcommand{\Rd}{R_{\rm d}}
\newcommand{\Gaia}{\textit{Gaia}}
\newcommand{\rcut}{r_{\rm cut}}
\newcommand{\rvir}{r_{\rm vir}}
\newcommand{\fdisc}{f_{\rm disc}}
\newcommand{\fhalo}{f_{\rm halo}}
\newcommand{\Pdisc}{P_{\rm disc}}
\newcommand{\Phalo}{P_{\rm halo}}
\newcommand{\hd}{h_{\rm d}}
\newcommand{\sigmarzero}{\sigma_{R,0}}
\newcommand{\Jcut}{J_{\rm cut}}
\newcommand{\Mh}{M_{\rm h}}

\newcommand{\vrad}{v_{\rm rad}}
\newcommand{\btheta}{\boldsymbol{\theta}}
\newcommand{\bXi}{\boldsymbol{\Xi}}

\newcommand{\Mtw}{M_{20}}
\newcommand{\Mvir}{M_{\rm vir}}
\newcommand{\logten}{\log_{10}}
\newcommand{\Rhole}{R_{\rm hole}}

\defcitealias{BW17}{BW17}

\begin{document}

   \title{Mass and shape of the Milky Way's dark matter halo with globular 
   clusters from \Gaia~and Hubble}

   \titlerunning{Mass \& shape of the Milky Way halo with clusters from \Gaia~\& HST}

   \author{Lorenzo Posti\inst{1}\fnmsep\thanks{posti@astro.rug.nl}
           \and
           Amina Helmi\inst{1}
          }

   \institute{Kapteyn Astronomical Institute, University of Groningen,
                          P.O. Box 800, 9700 AV Groningen, the Netherlands
%          \and
%              University of Alexandria, Department of Geography, ...\\
%              \email{c.ptolemy@hipparch.uheaven.space}
%              \thanks{The university of heaven temporarily does not
%                      accept e-mails}
             }

   \date{Received XXX; accepted YYY}

% \abstract{}{}{}{}{}
% 5 {} token are mandatory

  \abstract
   {}
   {We estimate the mass of the inner ($<20$ kpc) Milky Way and the axis ratio 
    of its inner dark matter halo
    using globular clusters as tracers. At the same time, we constrain the
    distribution in phase-space of the globular cluster system around the Galaxy.}
   {We use the \Gaia~Data Release 2 catalogue of 75 globular clusters' proper motions 
    and recent measurements of the proper motions of another 20 distant clusters obtained
    with the Hubble Space Telescope. We describe the globular cluster system with a distribution
    function (DF) with two components: a flat, rotating disc-like one and a rounder, more
    extended halo-like one. While fixing the Milky Way's disc and bulge, we let the mass and
    shape of the dark matter halo and we fit these two parameters, together with
     six others describing the DF, with a Bayesian method.}
   {We find the mass of the Galaxy within 20 kpc to be $M(<20{\,\rm kpc})=1.91^{+0.18}_{-0.17}
   \times 10^{11}\Msun$, of which $M_{\rm DM}(<20{\,\rm kpc})=1.37^{+0.18}_{-0.17}\times
   10^{11}\Msun$ is in dark matter, and the density axis ratio of the dark matter halo 
   to be $q=1.30 \pm 0.25$. 
   Assuming a concentration-mass relation, this implies a virial mass 
   $\Mvir = 1.3 \pm 0.3 \times 10^{12}\Msun$.
   Our analysis rules out oblate ($q<0.8$) and strongly prolate halos ($q>1.9$) with 
   99\% probability. Our preferred model reproduces well the observed phase-space
   distribution of globular clusters and has a disc component that closely resembles that of   
   the Galactic thick disc. The halo component follows a  power-law density profile 
   $\rho \propto r^{-3.3}$, has a mean rotational velocity of 
   $V_{\rm rot}\simeq -14\,\rm km\,s^{-1}$ at 20 kpc, and has a mildly radially biased
   velocity distribution ($\beta\simeq 0.2 \pm 0.07$, which varies significantly with 
   radius only within the inner 15 kpc). We also find that our distinction
   between disc and halo clusters resembles, although not fully, the observed distinction in metal-rich
   ([Fe/H]$>-0.8$) and metal-poor ([Fe/H]$\leq-0.8$) cluster populations.}
   {}


   \maketitle
%
%________________________________________________________________

\section{Introduction}

Giant leaps in the physical understanding of our Universe are often made 
when new superb datasets that peer into previously
uncharted territory become available. The second data release of
\Gaia~\citep{Gaia+18a} has just arrived and is by all means a perfect instance
of such a leap. The extent, accuracy, and quality of the data provided is
unprecedented, and it is leading the field of Galactic Astronomy to a completely
new era.

While  most of the new discoveries and exciting results are probably still hidden in
the vastness of the dataset, the increase in the number of objects observed and 
the greater accuracy, for example of the measured tangential motions on the sky, has the
potential to immediately lead to ground-breaking results \citep[][]{Gaia+18b,Gaia+18c,
Antoja+18}. In this study we exploit this novel dataset by employing a
well-established, but very sophisticated method to infer new tight constraints
on fundamental parameters, such as the total mass and the shape of the gravitational
potential of the inner Milky Way.

We do this by using the catalogue of absolute proper motions that \cite{Gaia+18b}
estimated for 75 globular clusters (GCs) around the Milky Way. Thanks to the unprecedented
performances of the \Gaia~astrometry, proper motions for these satellites could be
measured with a typical accuracy of a few tens of $\rm\mu as\,yr^{-1}$, which roughly
corresponds to an accuracy of $0.5-2\rm \,km\,s^{-1}$ in tangential velocity for a cluster
located at 10 kpc from us. Remarkable measurements of similar accuracy were also recently
released by \cite{Sohn+18} using the Hubble Space Telescope (HST). These authors estimated
proper motions of 20 clusters at large Galactocentric distances ($>10$ kpc),
using two HST epochs at least $\sim 6$ yr apart.

With the dataset provided by these two new catalogues, we can use the GCs as tracers of
the Galactic potential and hope to infer its total mass \citep[e.g.]
[]{BahcallTremaine81,Watkins+10}. 
Moreover, given the size and the unprecedented accuracy of the new catalogue, there may
well be enough signal in the data to constrain simultaneously the mass and the axis ratio
of the isodensity surfaces of the total mass distribution of the Galaxy,
similarly to previous work using the kinematics of individual stars in
the Milky Way halo \citep[e.g.][]{OllingMerrifield00,Smith+09,Bowden+16,Posti+17}.
In order to have enough inference from the data, we can model the Milky Way GC system
with equilibrium models in an arbitrary axisymmetric Galactic potential. One possibility
is then to use distribution functions (DFs) to represent the phase-space
distribution of the GC system and to measure the characteristic parameters of such DFs with
the data from \Gaia~and HST. This would not only allow the determination of the fundamental
parameters of the Galactic potential, but would also simultaneously constrain
the full phase-space distribution of the GCs around the Milky Way.

In this paper we use this approach and follow closely \citet[][hereafter
\citetalias{BW17}]{BW17}, who described the GC system with two distinct DFs, one with
halo-like dynamics (typically representing  the metal-poor clusters), and  one
with disc-like dynamics \citep[following the  more metal-rich clusters, see][]{Zinn85}.
These DFs are constructed in the space of the action integrals of motion. We build upon
the work of \citetalias{BW17} and improve on it in several ways: i) we use a much larger 
and more accurate dataset provided by very recent measurements presented above, ii)
we allow for the dark matter halo mass and shape to vary in our fit, and  iii) we simplify
the description of the DFs by fixing some characteristic parameters while still
reproducing remarkably well the observed distribution of GCs.

The paper is organised as follows: we introduce the data used in Section~\ref{sec:data}
and our modelling technique in Section~\ref{sec:model}; we describe our Bayesian
approach to determine the distribution of the model parameters in Section~\ref{sec:mcmc}
and we discuss our results in Section~\ref{sec:results};  finally, we  summarise and
conclude in Section~\ref{sec:concl}.
Throughout the paper we use a distance to the Galactic centre of $R_\odot=8.3$ kpc and
a peculiar motion of the Sun of $(U_\odot,V_\odot,W_\odot)=(11.1, 12.24, 7.25)\rm\,
km\,s^{-1}$ \citep{Schonrich+10}. We note, however, that reasonable differences in
$R_\odot$ and in the solar peculiar motion \citep[see \S3.2 and \S5.3.3 in][respectively]
{BlandHawthornGerhard16} are too small to significantly affect our conclusions.

\section{Data}\label{sec:data}

\subsection{Globular cluster data from Gaia DR2 and the Hubble Space Telescope}
Our dataset is primarily composed of the recently estimated proper motions and radial
velocities for 75 GCs from the \cite{Gaia+18b}. This catalogue
provides data of the best quality which is ideal to study the motion in phase-space
of GCs with unprecedented accuracy. The following observables are provided:
$(\alpha,\delta)$ are right ascension and declination, and $(\mu_{\alpha*},\mu_\delta)$
are the respective proper motions on the sky. Alongside these measurements, we
also have standard uncertainties and the correlation between the two proper motions
$C_{\mu_{\alpha*},\mu_\delta}$, which we use for generating samples from the error
distribution of each cluster. We supplement the \Gaia~DR2 catalogue with the
recent measurements of 16 other GCs from \cite{Sohn+18}. Hence, we work with an
unprecedented total of 91 GCs with exquisite proper motion data quality.

For both sets of clusters we compute heliocentric distances $s$ from the
extinction-corrected distance moduli in the $V$-band, taken from the latest
version of the catalogue compiled by \cite{Harris96}, and we assume the uncertainty
to be $0.05$ mag for each object. The heliocentric radial velocities $\vrad$ and their 
uncertainties are also taken from this catalogue. Therefore, the set of observables
that we use is
\begin{equation}
\bw = (\alpha,\delta,s,\mu_{\alpha*},\mu_\delta,\vrad)
,\end{equation}
where the error distribution of each cluster is assumed to be a multi-variate normal
distribution with non-null correlation $C_{\mu_{\alpha*},\mu_\delta}$ between the
proper motions and negligible variances in the sky positions $(\alpha,\delta)$ for
the \Gaia~DR2 clusters. We also add an additional systematic 35 $\rm\mu as\,yr^{-1}$
to the uncertainty budget of the cluster proper motion measured by \Gaia, as advised
in \cite{Gaia+18b}.

Finally, to make our tracer sample as complete as possible, we also consider 
52 GCs for which no  proper motion data is available. For this sample, we again obtain distance
moduli and radial velocities, together with their uncertainties  from \citet{Harris96}. 
This leaves us with a total sample of $91+52=143$~GCs.

\subsection{Sagittarius clusters}
\label{sec:sgr}

Several of the known GCs have been tentatively associated with the Sagittarius
dwarf galaxy/stream by different authors in the past \citep[see e.g.][]{LawMajewski10}. 
The Sagittarius dwarf galaxy is currently merging with our Galaxy \citep{Ibata+94}
and it is possibly bringing in a few GCs. If a significant fraction of the GCs that we use
as tracers are actually associated with the dwarf galaxy, this would imply a non-random
sampling of Galactic phase-space which could lead to biased estimates of the 
mass and shape of the Galactic potential using our DF-based method.

For this reason we run our algorithm  with the full sample of GCs, but also 
excluding the four clusters Arp 2, Palomar 12, Terzan 7, and Terzan 8 that
have been associated with  the Sagittarius dwarf galaxy according to the dynamical
analysis of \cite{LawMajewski10} and have been recently confirmed by \cite{Sohn+18}. 
We  retain M 54, the nuclear cluster in Sagittarius, as this
now traces the orbit of the dwarf singly. In addition, we also remove from our analysis
NGC 5053, which was recently found to be one of a pair of clusters
\citep[with NGC5024,][]{Gaia+18b}.

\section{Dynamical model}\label{sec:model}

Here we introduce the technique that we use to model the distribution in phase space
of the GC population given a Galactic gravitational potential. Our method borrows heavily
from \citetalias{BW17}, to which we refer the reader to for a more exhaustive description.
The Galactic potentials, the mapping from position-velocities
$(\bx,\bv)$ to action-angles $(\btheta,\bJ)$ and the DF models are constructed
with the \texttt{AGAMA} code \citep{Vasiliev18}.

%------ TAB. 1 --------------------------------------------------------------------------------
\begin{table}
\caption{Fixed parameters of the model \citep[see also][]{Piffl+14}}
\label{tab:pars}
\begin{center}
\begin{tabular}{lccc}
\hline\hline \\[-.2cm]
\multicolumn{4}{c}{Galactic Potential Parameters} \vspace{.1cm}\\
\hline\hline \\[-.2cm]
 & Thin disc & Thick disc & Gas disc \vspace{.1cm} \\
\hline \\[-.2cm]
$\Sigma_0/\Msun\,{\rm pc}^2$ &  $570.7$ & $251.0$ & $94.5$  \\
$\Rd/{\rm kpc}$ &  $2.7$ & $2.7$ & $5.4$ \\
$\hd/{\rm kpc}$ &  $0.2$ & $0.7$ & $0.04$ \\
$\Rhole/{\rm kpc}$ &  $0$ & $0$ & $4$ \vspace{.2cm} \\
\hline \\[-.2cm]
 & Bulge & & \vspace{.1cm} \\
 \hline \\[-.2cm]
$\rho_0/\Msun\,{\rm kpc}^3$ & $9.5\times 10^{10}$ & &  \\
$q$ & $0.5$ & & \\
$a$ & $0$ & &  \\
$b$ & $1.8$ & & \\
$r_0/{\rm kpc}$ & $0.075$  & & \\
$r_{\rm cut}/{\rm kpc}$ & $2.1$ & & \vspace{.2cm}\\
\hline\hline \\[-.2cm]
\multicolumn{4}{c}{Distribution Function Parameters}  \vspace{.1cm}\\
\hline\hline \\[-.2cm]
$R_{\sigma}/{\rm kpc}$ &  $13$ & & \\
$\hd/\Rd$ &  $0.2$ & & \\
$\Jcut/{\rm kpc\,km\,s^{-1}}$ & $10^5$ & & \\
$J_{\phi,0}/{\rm kpc\,km\,s^{-1}}$ & $100$ & & \\
\end{tabular}
\end{center}
\end{table}

\subsection{Galactic potential} \label{sec:pot}

We model the mass distribution of the Milky Way with five axisymmetric and analytic
components: a gaseous disc, stellar thin and thick discs, an oblate bulge, and a dark
halo. The discs are described by double-exponential density distributions
\begin{equation} \label{eq:doub_exp}
\rho(R,z) = \frac{\Sigma_0}{2\hd}\exp\left[-\frac{R}{\Rd}-\frac{|z|}{\hd}-\frac{\Rhole}{R}\right],
\end{equation}
where $\Rd$ and $\hd$ are scale-length and -height, $\Sigma_0$ is a normalisation
constant, and $\Rhole$ is the size of the central cavity (non-zero only for the gaseous disc), 
while the bulge and halo are described by
\begin{equation} \label{eq:halo}
\rho(R,z) = \frac{\rho_0}{m^a(1+m)^{b-a}}e^{-(m r_0/\rcut)^2},
\end{equation}
where $m=\sqrt{(R/r_0)^2+(z/qr_0)^2}$, $r_0$ is a scale radius, $q$ is the axis ratio, and
$\rho_0$ is a normalisation constant. The parameters of the three discs and the bulge
were fitted to a compilation  of observational constraints by \cite{Piffl+14}, most
notably to the kinematics of $\sim$ 200 000 stars in the solar neighbourhood as observed
by the RAdial Velocity Experiment \citep[RAVE,][]{Steinmetz+06}. These are kept fixed in
our analysis and are summarised in Table~\ref{tab:pars}.

For  the dark matter halo instead, we make a cosmologically motivated ansatz fixing the
slopes in Eq.~\eqref{eq:halo} to those of a \citet[][hereafter NFW, $a=1$ and $b=3$]
{NFW96} model, truncating the halo with an exponential cut-off at the virial radius,
$\rcut=\rvir$, and imposing the concentration-mass relation found in $\Lambda$CDM N-body
simulations: $\logten\,c = 1.025 - 0.097\logten(M/10^{12}h^{-1}\Msun)$, where $c=\rvir/r_0$ is
the concentration and $h=0.68$ is the Hubble parameter \citep[see][]
{DuttonMaccio14}. Thus, our halo model is described by two free parameters: a mass and
the axis ratio $q$.

\subsection{Distribution function of the GC population}

Assuming the GCs to be in dynamical equilibrium within the potential of the Galaxy, we
can describe their distribution in phase-space with a function $f(\bJ)$ of the three
action integrals $\bJ=(\Jr,\Jphi,\Jz)$. This function, the DF, is normalised such that
$(2\pi)^3 f(\bJ)\,\de\bJ$ is the probability that a GC moves on the orbit specified by
the action triplet $\bJ$. In a general axisymmetric potential, the action integrals can
be computed from positions and velocities, $\bJ=\bJ(\bx,\bv)$, with the `Staeckel Fudge'
algorithm \citep{Binney12}.

Following \cite{Zinn85}, who suggested that the GC system is composed of two populations
distinct in their metallicity and their phase-space distribution, we allow the DF to have two
dynamically distinct components, one disc-like and one halo-like: 
$f(\bJ)=\fdisc(\bJ)+\fhalo(\bJ)$.
For the disc component, we use the `quasi-isothermal' DF, introduced by \cite{Binney10},
as implemented in \cite{Vasiliev18}
\begin{equation} \label{eq:discDF}
\fdisc(\bJ) = \frac{\Sigma\nu\Omega}{2\pi^2\kappa\sigmar^2\sigmaz^2}
  \exp\left(-\frac{\kappa\Jr}{\sigmar^2}-\frac{\nu\Jz}{\sigmaz^2}\right)
  f_{\pm,\rm d},
\end{equation}
where $\Omega,\kappa$, and $\nu$ are the circular, radial, and vertical epicycle
frequencies evaluated at the radius of the circular orbit $\Rc=\Rc(\tilde{J})$ with
angular momentum $\tilde{J}=|\Jphi|+\Jr+\Jz$; the factor controlling the disc surface
density is $\Sigma=\Sigma_0 \exp[-\Rc(\Jphi)/\Rd]$ and
\begin{equation} \label{eq:frot_disc}
f_{\pm,\rm d} = \begin{cases}
                                1 & \Jphi\geq0, \\
                                \exp(2\Omega\Jphi/\sigmar^2) & \Jphi<0,
                            \end{cases}
\end{equation}
makes the DF odd in the angular momentum $\Jphi$ and controls the net rotation of the
disc component. The factors $\sigmar$ and $\sigmaz$, which determine the radial and
vertical velocity dispersions, are chosen to mimic the Galactic thick disc model 
as in \citet{Piffl+14}, with an
exponential vertical velocity dispersion with constant scale-height,
$\sigmaz=\sqrt{2}\hd\nu$, and a radial velocity dispersion
$\sigmar= \sigmarzero\exp(-\Rc/R_{\sigma})$.
In analogy with the thick disc, we fix $R_{\sigma}=13$ kpc \citep[as measured by][]
{Piffl+14} and $\hd = 0.2 \Rd$. We are thus left with two free parameters to be fitted,
the disc scale-length $\Rd$ and the central radial dispersion $\sigmarzero$, that
completely characterise $\fdisc(\bJ)$. 

For the halo component, we use the double power-law DF, introduced by \citet[]
[see also \citeauthor{WilliamsEvans15} \citeyear{WilliamsEvans15}]{Posti+15}, as
implemented in the \texttt{AGAMA} code. Here we fix the halo DF to have a constant
density core in phase-space, thus making the 3D density distribution of the model
a single power law with a central core; the halo DF then is
\begin{equation}\label{eq:haloDF}
\fhalo(\bJ)=\frac{\Mh}{(2\pi J_0)^3}[1+g(\bJ)/J_0]^{-B} \exp[-(g(\bJ)/\Jcut)^2]
  f_{\pm,\rm h},
\end{equation}
where $J_0$ is a scale action defining the size of the constant density core,
$\Jcut$ is a (large) cut-off action, $g(\bJ)=k_R\Jr+(3-k_R)(|\Jphi|+\Jz)/2$ is
a homogeneous function of degree one and
\begin{equation} \label{eq:frot_halo}
f_{\pm,\rm h} = 1 + \chi\tanh(\Jphi/J_{\phi,0})
\end{equation}
is the factor controlling the net rotation of the system, with $\chi=0$ being the
non-rotating case and $\chi=\pm 1$ being the maximally rotating/counter-rotating case.
Here we fix  two parameters: i) the cut-off action to $\Jcut=10^5$ 
kpc km/s, so that the halo density distribution is cut off at distances much larger
than that of the farthest GC ($\gtrsim 300$ kpc) ensuring a finite halo mass, and ii)
$J_{\phi,0}=100$ kpc km/s, 
so that the possible rotation of the halo smoothly goes to zero to the centre
\citepalias[needed for continuity of the
DF, see][]{BW17}. These two are the only nuisance parameters of the halo DF, and
we are left with four free parameters to be fitted: the slope $B$,
the extent of the constant density core $J_0$, the rotation parameter $\chi$, and
$k_R$ which controls the system's velocity anisotropy. 

Our choice for the GC halo DF is relatively simple, describing the density
profile of the halo GCs with a single power law. This is convenient because the
small number of GCs at distances $>20$ kpc limits our inference on radial trends. 
Therefore, it does not have the same level of complexity as seen in the stellar halo
of the Galaxy \citep[e.g.][]{Carollo+07,Deason+11,Sesar+13,Xue+15,DasBinney16,Iorio+18}.

Finally, we note that the normalisation parameters $\Sigma_0$ for $\fdisc$ and $\Mh$
for $\fhalo$ are unimportant in our case, since the GC system is treated as mass-less
and simply traces the gravitational potential.

\section{Parameter estimation} \label{sec:mcmc}

We estimate the posterior distribution of the model parameters $\bXi$ that best
represent the data $\bw$ using standard Bayesian inference
\begin{equation}\label{eq:bayes}
P(\bXi|\bw) = \frac{P(\bw|\bXi)\,P(\bXi)}{P(\bw)},
\end{equation}
where $P(\bXi)$ is the prior, $P(\bw|\bXi)$ is the likelihood, and $P(\bw)$ is the
evidence, which is unimportant for our purposes and hence is neglected.

Our model has eight free parameters. There are two  for the disc DF: the disc scale-length $\Rd$ and
the central velocity dispersion $\sigmarzero)$;  four for the halo DF: the slope $B$,
scale action $J_0$, anisotropy coefficient $k_R$, and rotation parameter $\chi$; and two for the dark matter halo potential: the axis ratio $q$ of the density
distribution and, instead of varying $\Mh$ in Eq.~\ref{eq:haloDF}, we use the 
the total mass of the Galaxy within 20 kpc, which is a characteristic radius probed
by the tracers, $\Mtw =M_{20,\rm baryons}+M_{20,\rm DM}$, where 
$M_{20, \rm baryons}=5.4\times 10^{10}\Msun$ is a constant that is fixed by the 
baryonic model in Sect.~\ref{sec:pot}.
Following \cite{Bowden+16},  we transform the axis ratio to the quantity
\begin{equation}\label{eq:u}
u=\frac{2}{\pi}\arctan(q),
\end{equation}
which varies in a finite range $[0,1]$. A spherical model has $u=1/2$, oblate halos
have $0<u<0.5$, while prolate ones have $0.5<u<1$.

\subsection{Prior}

We adopt non-informative priors for all parameters. Three of these have finite domains,
hence we use uniform priors for them:
\begin{equation}\label{eq:pr_uni}
0<k_R<3, \qquad -1\leq \chi \leq 1, \qquad 0<u<1.
\end{equation}
The other five are intrinsically positive quantities, hence non-informative priors
are uniform in the logarithm:
\begin{eqnarray}\label{eq:pr_log}
-1\leq \logten\Rd/{\rm kpc}\leq 3,& \qquad
  0\leq \logten\sigmarzero/{\rm km\,s^{-1}} \leq 4, \nonumber\\
0\leq \logten B\leq 2,&   \qquad 0\leq \logten J_0 /{\rm kpc\,km\,s^{-1}}\leq 4, \\
8\leq \logten\Mtw/\Msun\leq 13.\nonumber
\end{eqnarray}
Thus, our prior $P(\bXi)$ is the product of these eight terms.

\subsection{Likelihood}

The likelihood $P(\bw|\bXi)$ that the $N$ globular clusters in our dataset are  moving in the
gravitational potential $\Phi$ and are drawn from the phase space distribution
$f(\bJ|\Phi)$ is
\begin{equation} \label{eq:likelihood}
P(\bw|\bXi) = \prod_{j=0}^N P(\bw_j|\bXi) = \prod_{j=0}^N \int\de\bw\,
        G_j(\bw,\overline{\bw}_j,\mathsf{C}_j)f(\bJ|\Phi)
    \left|\frac{\partial(\btheta,\bJ)}{\partial\bw}\right|,
\end{equation}
where the Jacobian factor is \citepalias[see e.g.][]{BW17}
\begin{equation}\label{eq:jac}
\left|\frac{\partial(\btheta,\bJ)}{\partial\bw}\right| = s^6\cos b.
\end{equation}
In Eq.~\eqref{eq:likelihood} we have convolved the DF of the model with the
error distribution $G_j$ of each cluster. $G_j$ is a multi-variate
normal with mean $\overline{\bw}$ given by the measurements and covariance
matrix $\mathsf{C}$, whose only non-null off-diagonal element is given by
the correlation coefficient $C_{\mu_{\alpha*},\mu_\delta}$, which is
measured only for the 75 clusters analysed with \Gaia~DR2. To compute the
integral in Eq.~\eqref{eq:likelihood} we use a similar strategy to
that of \citetalias{BW17}: we employ an importance-sampling Monte Carlo method,
with the same sampling density as in \citetalias{BW17} (their $f_{\rm S}$
as in their Section 3.3.3).

In this work we assume that our sample of GCs is complete \citep[e.g.][]
{Harris01}, meaning that we neglect the possibility that a significant number of
GCs is still hidden by dust close to the Galactic mid-plane. 
This approach is validated by \citetalias{BW17}, who have shown
that including a selection function depending on dust extinction (as an extra
factor multiplying the integrand in Eq.~\ref{eq:likelihood}) does not alter
the final inference on the model parameters. 

%------ FIG. 1 --------------------------------------------------------------------------------
\begin{figure*}
\includegraphics[width=\textwidth]{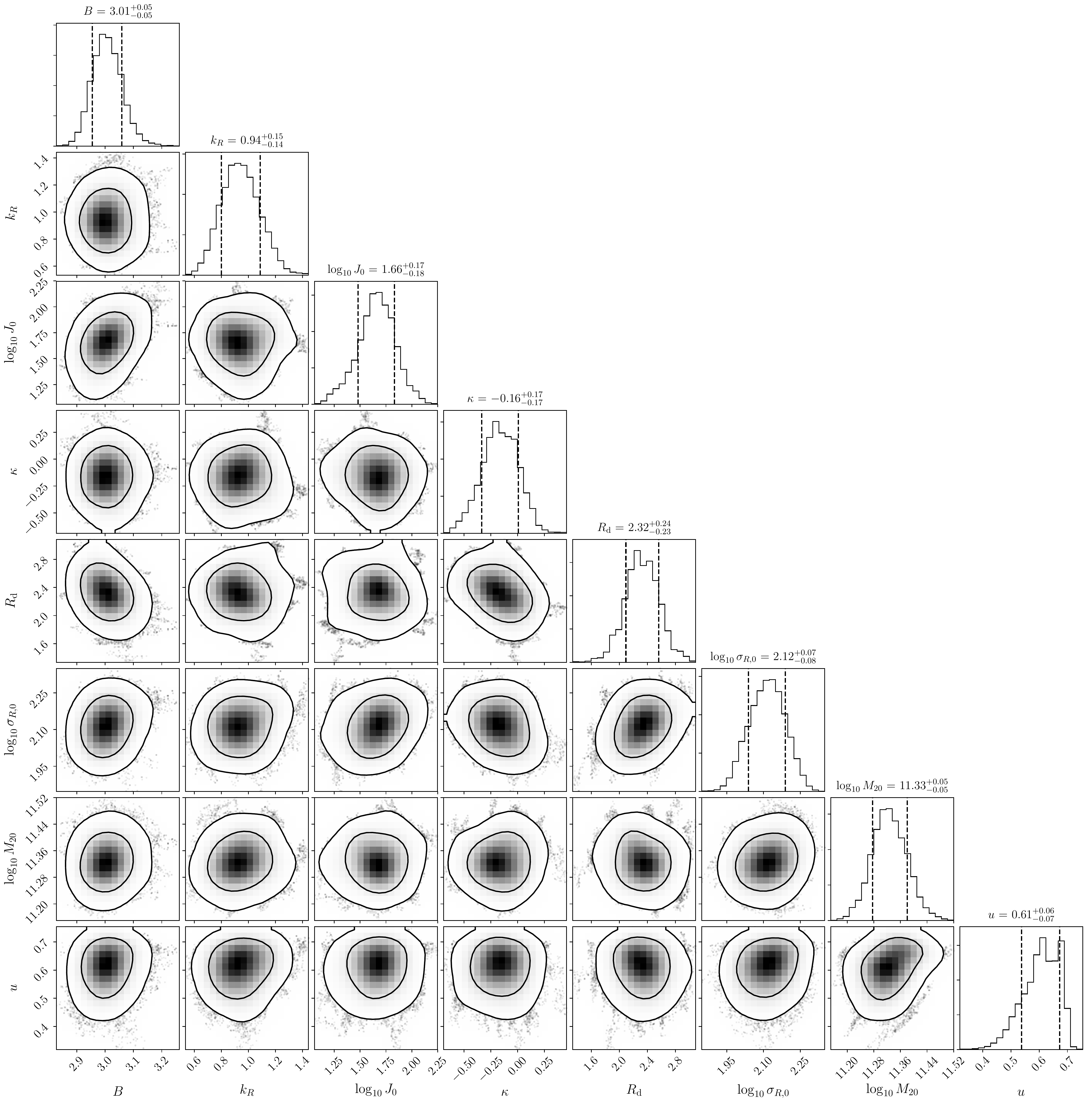}
\caption{Posterior distributions for the eight model parameters. The black solid curves
         in the 2D marginalised subspaces enclose $68\%$ and $95\%$ of the total 
         probability. The black vertical dashed lines in the 1D marginalised subspaces
         are the 16th and 84th percentiles.  
        }
\label{fig:mcmc_full}
\end{figure*}

\subsection{Posterior}

We sampled the posterior distribution $P(\bXi|\bw)$ of the model parameters via
a Markov chain Monte Carlo (MCMC) method. We used a Metropolis-Hastings sampler
\citep[][]{Hastings70} with a multi-variate proposal distribution.
We ran 50 independent chains for 3000 steps, which were found to converge after
a short burn-in phase of about $\sim 600$ steps, thus we effectively sampled the
posterior distribution with $\sim 120,000$ samples.
All the chains turned out to be well mixed and we tuned the
parameters of the proposal distribution to have $\sim 40-60\%$ acceptance
rate and small auto-correlations (after burn-in).

\section{Results}\label{sec:results}

Figure \ref{fig:mcmc_full} shows the  1D and 2D marginalised posterior distributions of
the eight free parameters of our model. All the parameters are relatively well constrained
by our analysis, with relative uncertainties limited to $\sim 10\%$. 
In what follows we  always associate error bars with each measurement
representing the 16th and 84th percentiles of the marginalised posterior distribution.
Figure \ref{fig:mcmc_full} depicts some non-zero correlations between the parameters which we discuss below.

\begin{figure}
\includegraphics[width=0.49\textwidth]{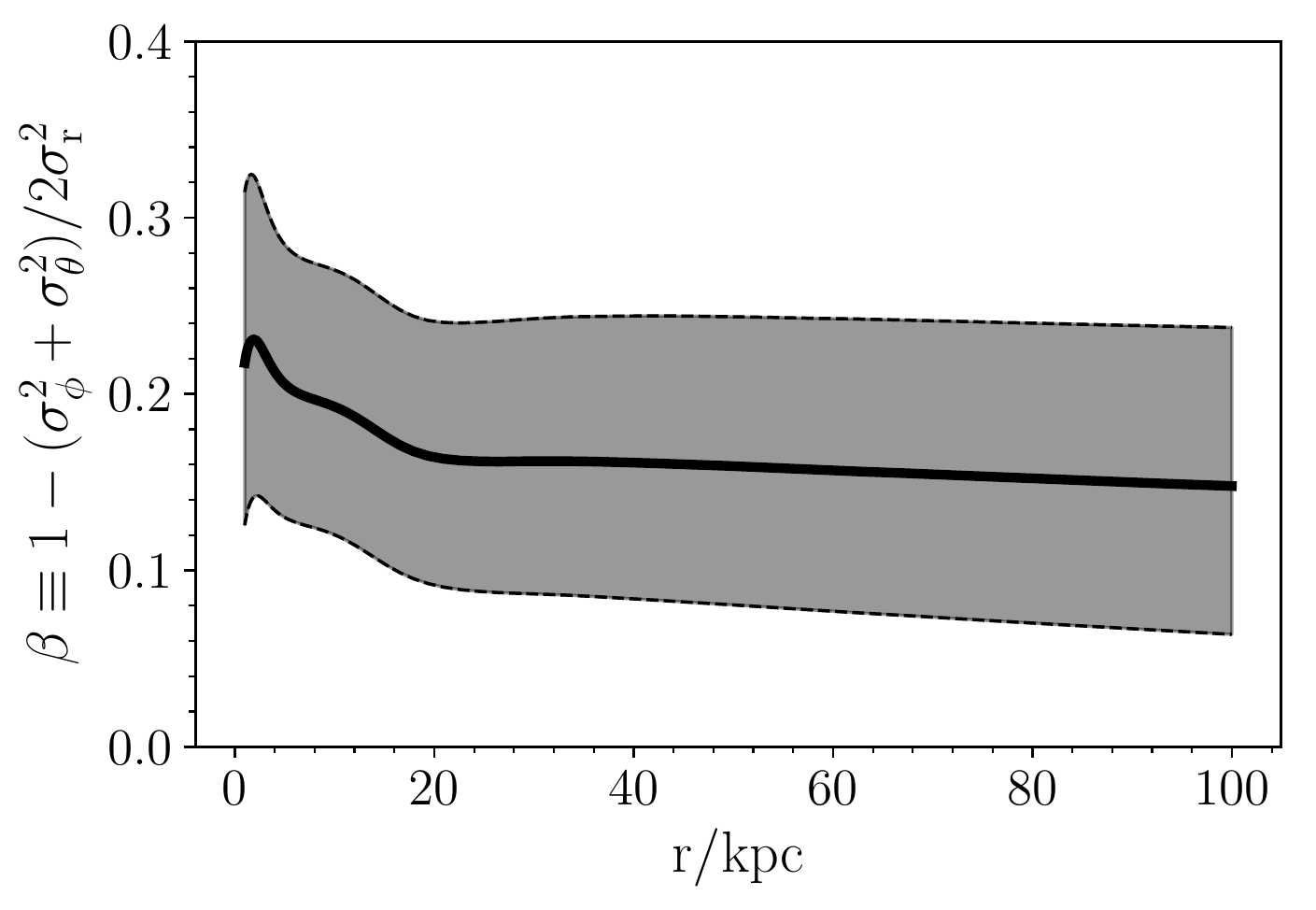}
\caption{Spherically averaged velocity anisotropy profile of the halo component of the
                 GC system. The black solid line is for our best model and the grey band encompasses
         the 16th and 84th percentiles of the distribution.
        }
\label{fig:halo_anisotropy}
\end{figure}

\subsection{Distribution function}

\subsubsection{Halo}
The halo DF of the best model has a shallow slope ($B=3.01\pm 0.05$) and a
small-scale action ($J_0=46^{+22}_{-15}$ kpc km s$^{-1}$) and we find these two
quantities to be correlated as models with a steeper slope have a larger scale
action. This degeneracy is not surprising, and it is inevitable since $J_0$ controls
the physical scale at which the spatial density profile steepens. Our best model
has a very small constant density core, of  $\sim 0.1$ kpc, and its density
distribution is a power law close to $\rho\propto r^{-3.3}$.

Contrary to previous claims \citepalias[e.g.][]{BW17}, we find the halo component
not to be significantly rotating ($\kappa=-0.14\pm 0.14$), if not mildly
counter-rotating. In fact, in our best model the halo component has a small net
retrograde rotation of the order of $V_{\rm rot}\sim -14$~km/s at 20 kpc
\citep[roughly consistent with other independent estimates, e.g.][]{Beers+12,
Kafle+17,Koppelman+18}, while the disc component is rotating at about
$V_{\rm rot}\sim 210$~km/s at the solar radius. 
This difference compared to previous work is due to
the dramatic increase in data quality.

We also constrain relatively well the coefficient $k_R$ of the homogeneous function
$g(\bJ)$ entering in the halo DF definition, and which controls the velocity anisotropy
of the halo component. Our best model with $k_R=0.94^{+0.15}_{-0.14}$ has a mildly
radially biased velocity distribution, with a rather constant $\beta\equiv 1 -
(\sigmaphi^2+\sigma_\phi^2)/2\sigma_r^2\simeq 0.15-0.2$, consistent
with previous measurements \citep[e.g.][]{Eadie+17}. We show in Figure
\ref{fig:halo_anisotropy} the spherically averaged anisotropy profile of the halo
component.

\subsubsection{Disc}

The disc DF is consistent with that of the Galactic thick disc fitted by
\cite{Piffl+14}: the scale-length of the disc is $\Rd=2.32^{+0.24}_{-0.23}$ kpc
(see Table~\ref{tab:pars}),
while the central radial velocity dispersion is $\sigmarzero=132^{+23}_{-22}$ km/s.
However, given that we have fixed two of the DF parameters, $R_{\sigma}$
and $\hd$, to resemble the thick disc, it comes as no surprise that we find
agreement with the estimates by \cite{Piffl+14}. This serves more as a sanity check
that the MCMC chains are converging to a physical solution, instead of a proper
constraint. The fact that we find the most likely models precisely in the region
of the parameter space where the Galactic thick disc is, implies that our
analysis is consistent with the possibility that some GCs were born in the
thick disc or were associated with its formation \citep[e.g.][]{Zinn85}.

\subsubsection{Spatial and velocity distribution}

\begin{figure}
\includegraphics[width=0.49\textwidth]{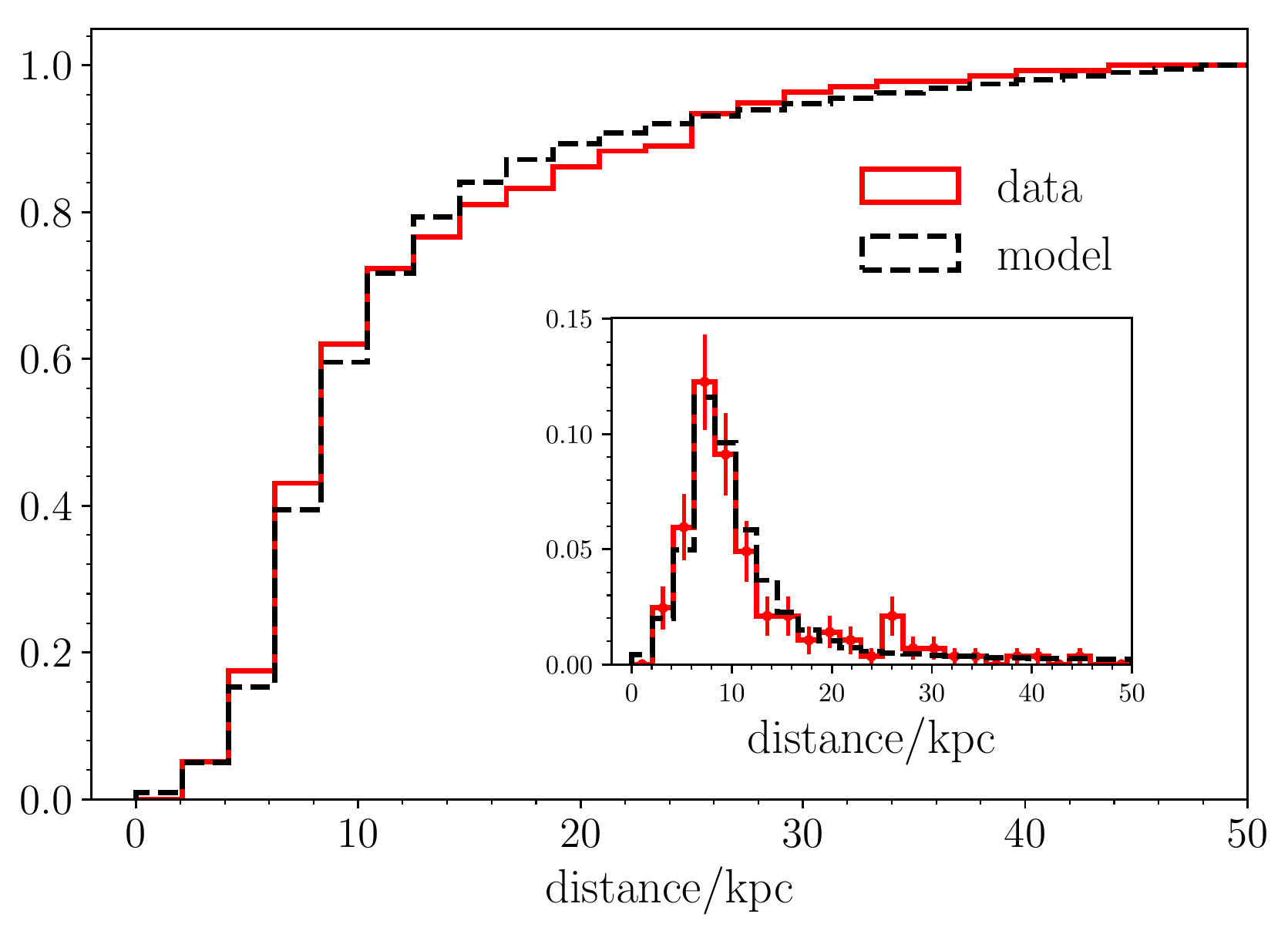}
\caption{Cumulative distribution of heliocentric distances of the 143 GCs (red solid)
         compared to that of the best model (black dashed). We cut these at $50$ kpc
         for illustration purposes. The Kolmogorov-Smirnov test returns a
         $63\%$ probability that the two samples are drawn from the same distribution.
         The inset shows the normalised distribution of distances within $30$ kpc
         with Poissonian error bars.
        }
\label{fig:hist_distance}
\end{figure}

We now compare the observed and predicted spatial and velocity distributions of
the GC system. In Figure \ref{fig:hist_distance} we show the heliocentric distance
distribution of the full sample of 143 GCs compared to that of our best model.
For visualisation purposes, we have cut both distributions at $50$ kpc.
The model represents very well the observed distances to the cluster: running a
Kolmogorov-Smirnov test on the full sample (143 GCs)  we find a $63^{+15}_{-10}\%$
probability\footnote{
Uncertainties are estimated with 1000 bootstrap realisations.
} that they are drawn from the same distribution.
It turns out that most of the numerical noise in this comparison comes from the
six clusters found at distances between $50$ and $120$ kpc: if we exclude this large and
very poorly sampled volume and we restrict the comparison to the clusters
within 50 kpc we find a probability of $93^{+5}_{-4}\%$ that the data and model
come from the same underlying distribution. Within 50 kpc, the largest discrepancy
is found between 10 and 30 kpc, but it is not significant (see  inset in
Fig.~\ref{fig:hist_distance}).

\begin{figure*}
\includegraphics[width=1\textwidth]{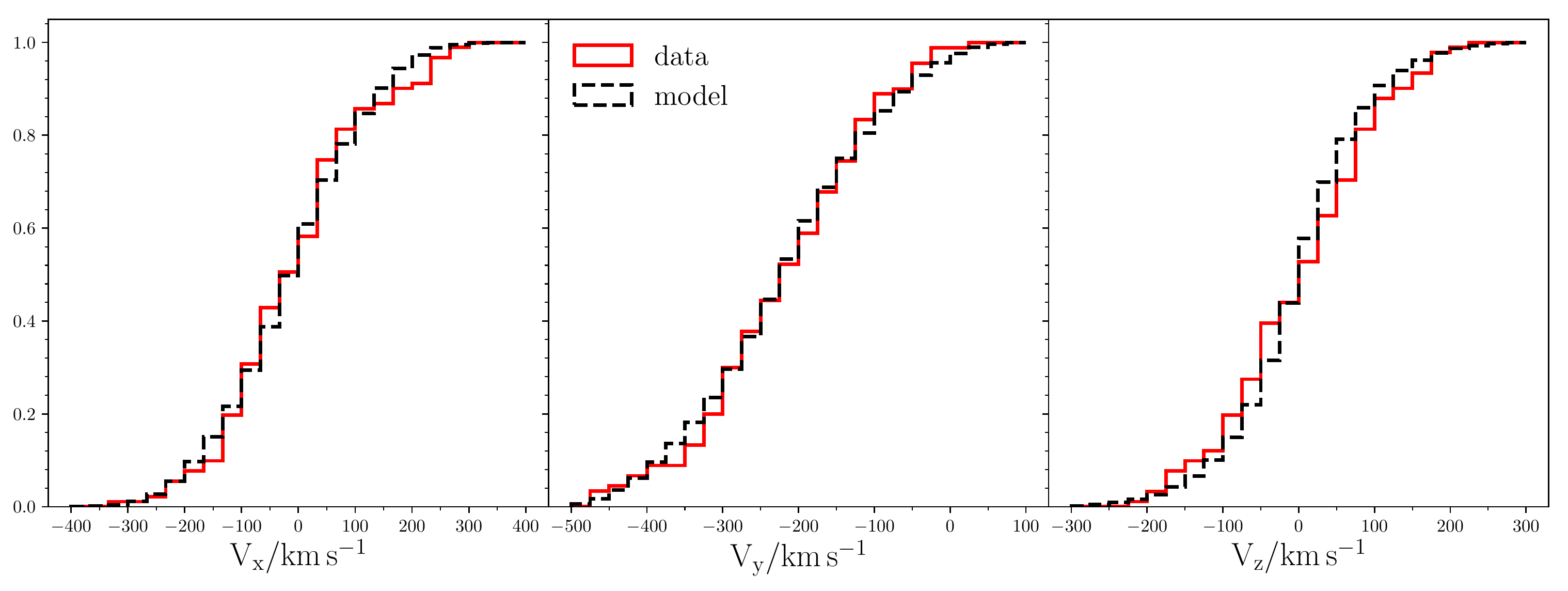}
\caption{Cumulative distribution of heliocentric cartesian velocities of the GCs
         (red solid) compared to that of the best model (black dashed).
         The Kolmogorov-Smirnov test returns a
         $88\%$, $83\%$, and $42\%$ probability that the two samples in $V_x$, $V_y$,
         and $V_z$, respectively, are drawn from the same distributions. 
        }
\label{fig:vel_distrib}
\end{figure*}

\begin{figure}
\includegraphics[width=0.49\textwidth]{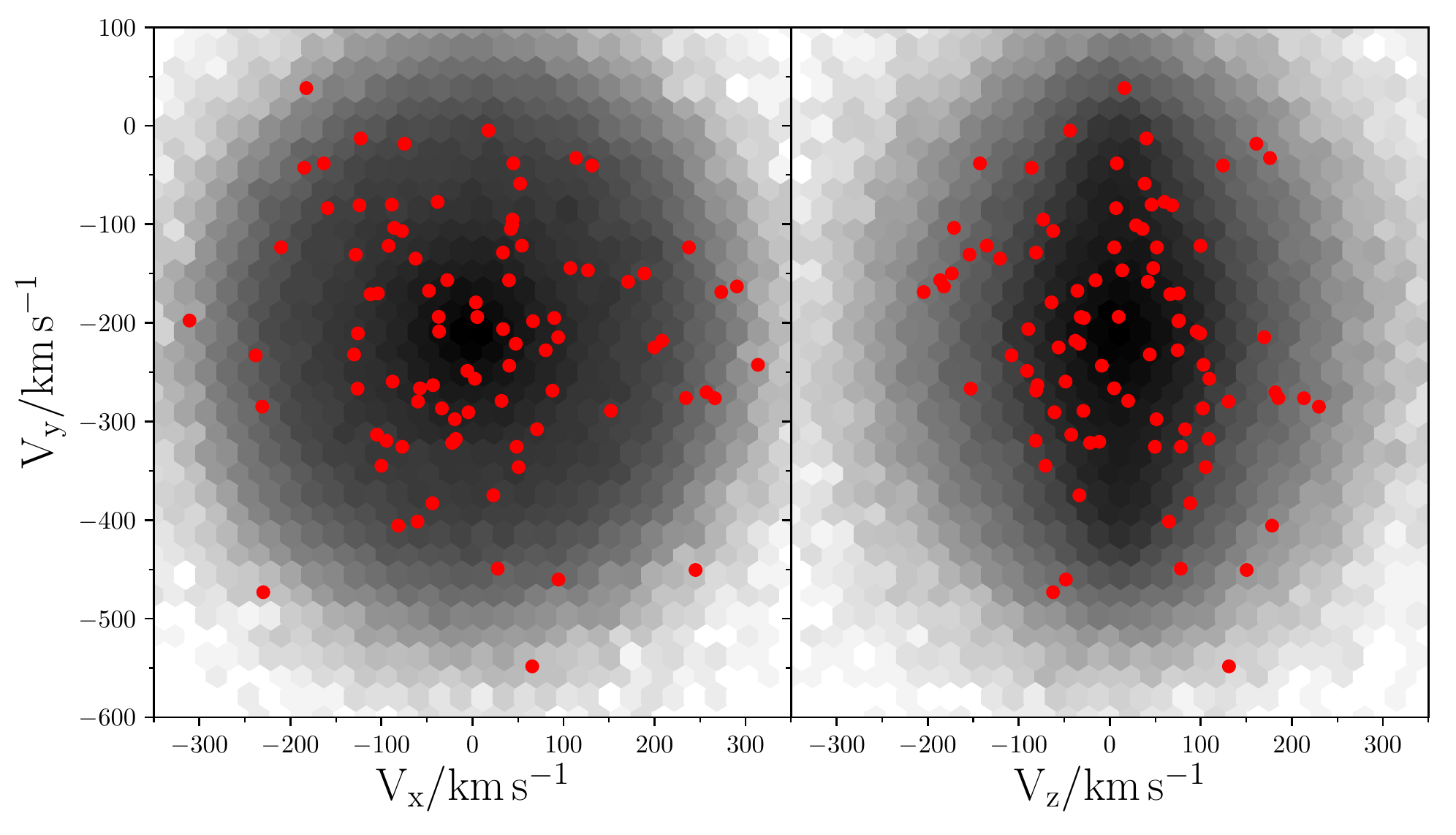}
\caption{Velocity distribution in the $V_x-V_y$ and $V_z-V_y$ planes of the model
                 (black) compared to that of the data (red circles).
         }
\label{fig:2d_vel_distrib}
\end{figure}

For the first time we have a sufficiently large dataset of accurate proper
motions of 91 GCs, and  we can also make a meaningful comparison  with the velocity
distribution predicted by the model. We show in Figure \ref{fig:vel_distrib} the
cumulative distributions of three cartesian heliocentric velocities $(V_x,V_y,V_z)$,
where $V_x$ points towards the Galactic centre and $V_Y$ is parallel to the Galactic
disc's rotation. 
Figure \ref{fig:2d_vel_distrib} also shows  the 2D projections of
the clusters velocity distribution compared to the models.

The agreement between data and model is generally very good; however, it should
be noted that the model appears somewhat more symmetric with respect to the observed
velocity distribution of GCs. On the one hand, it should be kept in mind that our equilibrium
models will always produce symmetric distributions in $V_x$ and $V_z$; on the other hand,
observed deviations from symmetry may be due to the presence of substructures
or may simply be stochastic, due to the small number of  tracers (91).

\subsubsection{Metallicity distribution}

For each model we can use Eq.~\eqref{eq:likelihood} to compute the probability
that a cluster belongs to the disc component rather than to the halo
by substituting $f$ with $\fdisc$ or $\fhalo$, respectively. The ratio of these
two probabilities, which we call $\Pdisc$ and $\Phalo$, can be used to estimate
the membership of each GC to these components. We now compare these membership 
probabilities with
the metallicity [Fe/H] of each cluster, whose distribution is bimodal with a
minimum at [Fe/H]$\simeq-0.8$ that has been classically used to distinguish
metal-poor halo clusters from metal-rich disc ones \citep[e.g.][]{Zinn85}.

\begin{figure}
\includegraphics[width=0.49\textwidth]{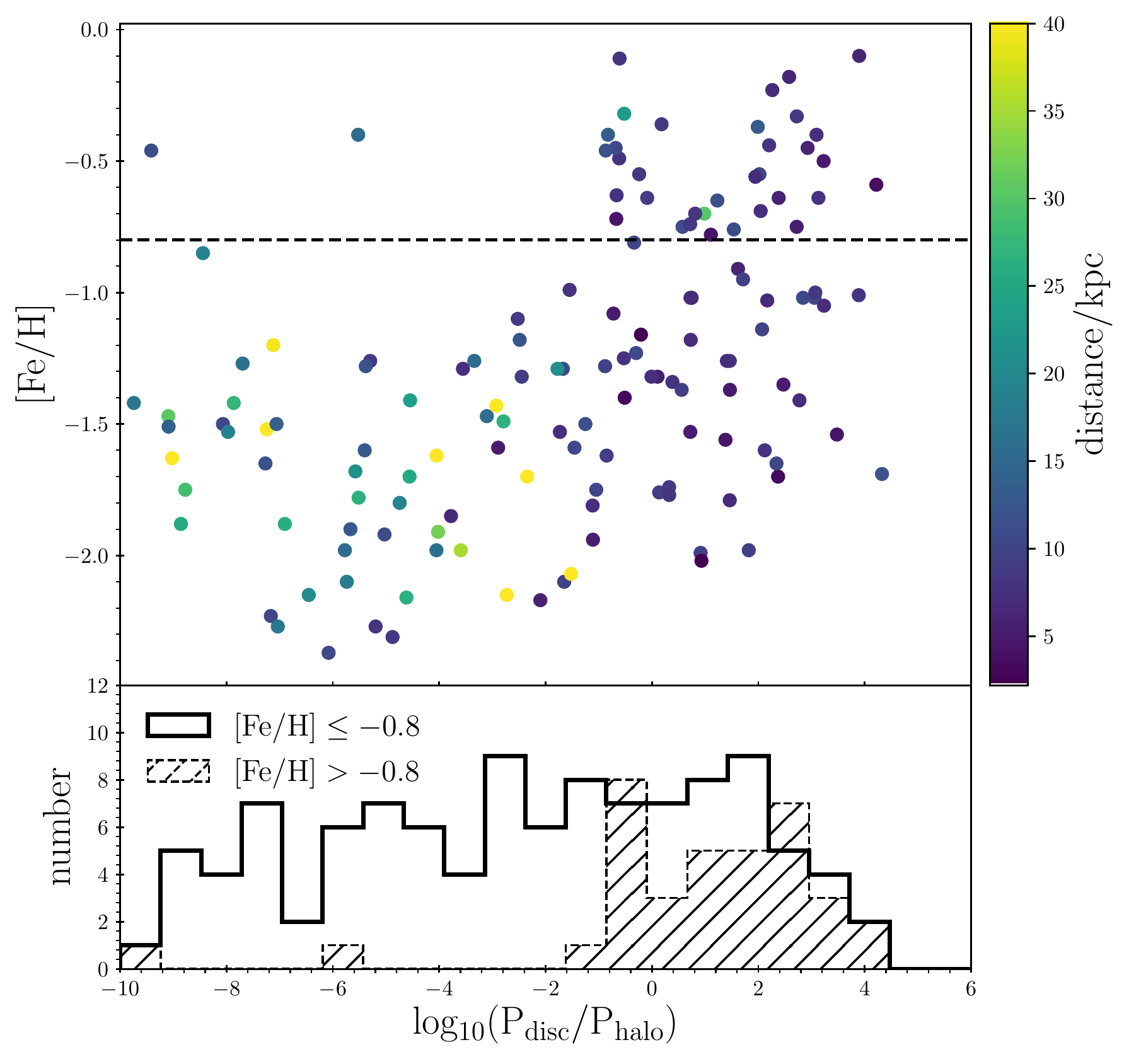}\\
\caption{\emph{Upper panel:} Metallicity of the GCs as a function of the ratio of the
         probability of belonging to the disc or the halo component of the DF.
         The points are colour-coded according to the heliocentric distance of the
         cluster. The horizontal dashed line marks [Fe/H]=-0.8, the classical distinction
         between metal-poor and metal-rich clusters.
         \emph{Lower panel:} Histogram of the ratio of probabilities in two bins of
         metal-rich ([Fe/H]$>$-0.8) and metal-poor ([Fe/H]$\leq$-0.8) clusters.
        }
\label{fig:met_prob}
\end{figure}

Figure \ref{fig:met_prob} shows the metallicity of the 143 GCs as a function
of the logarithm of the ratio $\Pdisc/\Phalo$ in our best model. Similarly to
\citetalias{BW17}, we find that metal-rich clusters are typically much more
likely to belong to the disc component rather than the halo, with only a couple
of exceptions. In particular, all but two clusters with [Fe/H]$>-0.8$ have
$\logten(\Pdisc/\Phalo)>-2$, making it  very unlikely that they are a part of the halo component
(see  histogram in the bottom panel of Fig.~\ref{fig:met_prob}). At lower
metallicities, instead, there are many more clusters with $\logten(\Pdisc/\Phalo)
<-2$ ($67\%$ below [Fe/H]$\leq-0.8$), making them very likely halo clusters.
These are typically objects that are either found at very large distances,
that are counter-rotating, or that have negligible angular momentum. 
Figure~\ref{fig:met_prob} shows that only a few clusters with heliocentric
distances larger than 20 kpc have a higher probability of being part of the
disc component.

\begin{figure*}
\includegraphics[width=\textwidth]{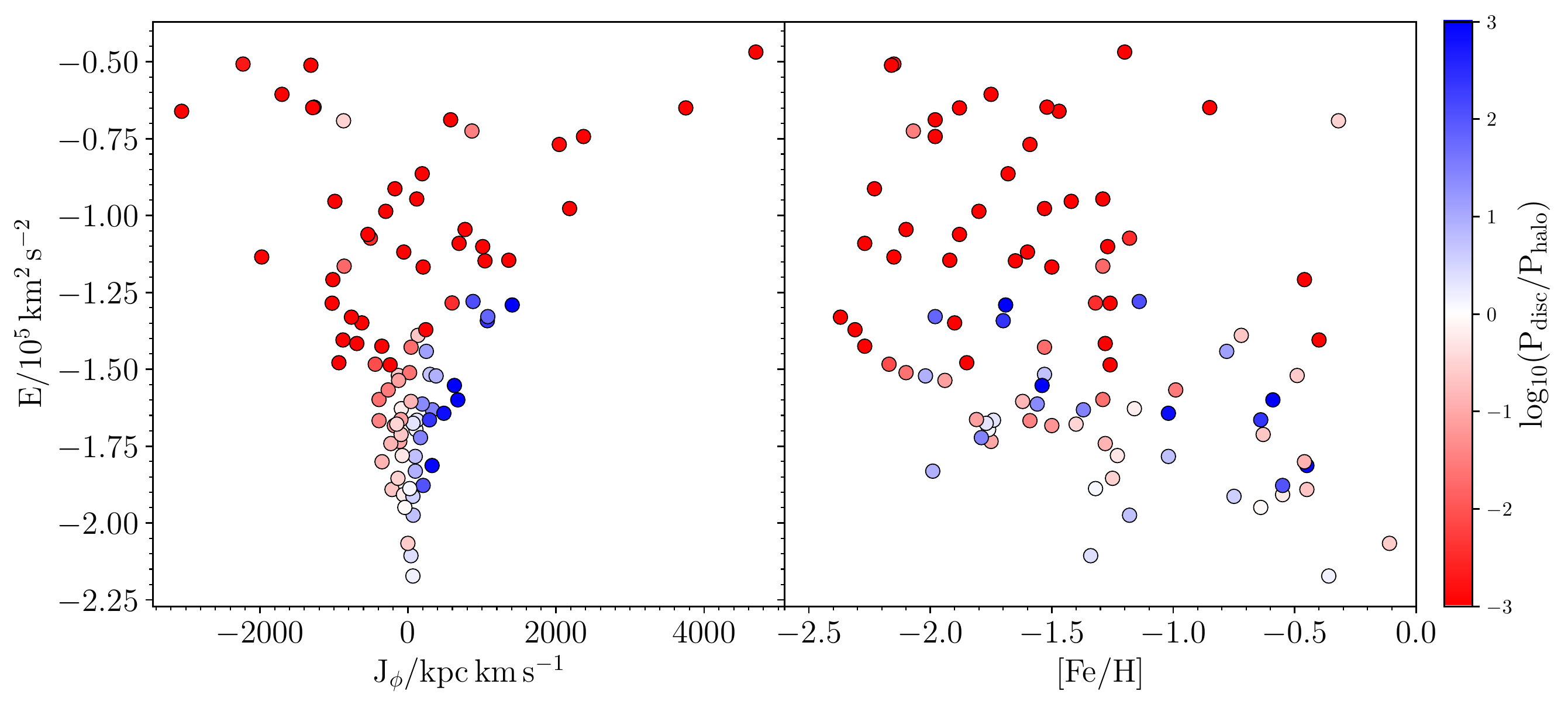}\\
\caption{\emph{Left panel:} Distribution of the 91 GCs with accurate
         proper motions in the angular momentum-energy plane, 
         colour-coded by the ratio of probability
         of belonging to the disc or to the halo.
         \emph{Right panel:} Same as the left panel, but in the
         metallicity-energy plane.
        }
\label{fig:met_ELz}
\end{figure*}

In the left-hand panel of Figure \ref{fig:met_ELz} we show the
distribution of the 91 clusters with accurate proper motions in the plane 
of integrals of motion angular momentum-energy, colour-coded by 
$\logten(\Pdisc/\Phalo)$. Here we see that all  clusters  less bound than 
$E\gtrsim -1.25 \times 10^5\,\rm km\,s^{-1}$ belong to the halo component.
At lower (more negative) energies, halo clusters clearly prefer retrograde ($\Jphi<0$)
or non-rotating ($\Jphi\sim 0$) orbits, while the probability of belonging
to the disc component is maximal for prograde angular momenta. For the likely 
disc clusters ($\logten\Pdisc/\Phalo>0$) we find
an average angular momentum of 
$\overline{J}_\phi = 364\pm 71 \,\rm kpc\,km\,s^{-1}$, while for the very likely
halo clusters ($\logten\Pdisc/\Phalo<0$) we find  
$\overline{J}_\phi = -164\pm 112 \,\rm kpc\,km\,s^{-1}$, from which we conclude that
the halo clusters exhibit a hint of retrograde rotation. 
If we divide the sample of 91 GCs according to the traditional separation into
metal-rich [Fe/H]$>-0.8$ (18 clusters) and metal-poor [Fe/H]$\leq-0.8$ 
(73 clusters), we find a mean $z$-angular momentum of respectively
$\overline{J}_\phi=-70\pm 120 \,\rm kpc\,km\,s^{-1}$ and 
$\overline{J}_\phi=17\pm 100\,\rm kpc\,km\,s^{-1}$ for the two populations.
This is because some metal-rich clusters are on low-energy, sometimes even
retrograde orbits, while a few clusters on disc-like orbits happen to have
low  metallicity.

The right-hand panel of Figure \ref{fig:met_ELz} shows energy as a
function of metallicity, with the points being again colour-coded by
their membership probability. It can be seen that the metal-poor halo
clusters are mostly confined to less bound orbits, while 
disc clusters with high angular momentum span a wide range of metallicities.

\subsection{Gravitational potential}

\begin{figure}
\includegraphics[width=0.49\textwidth]{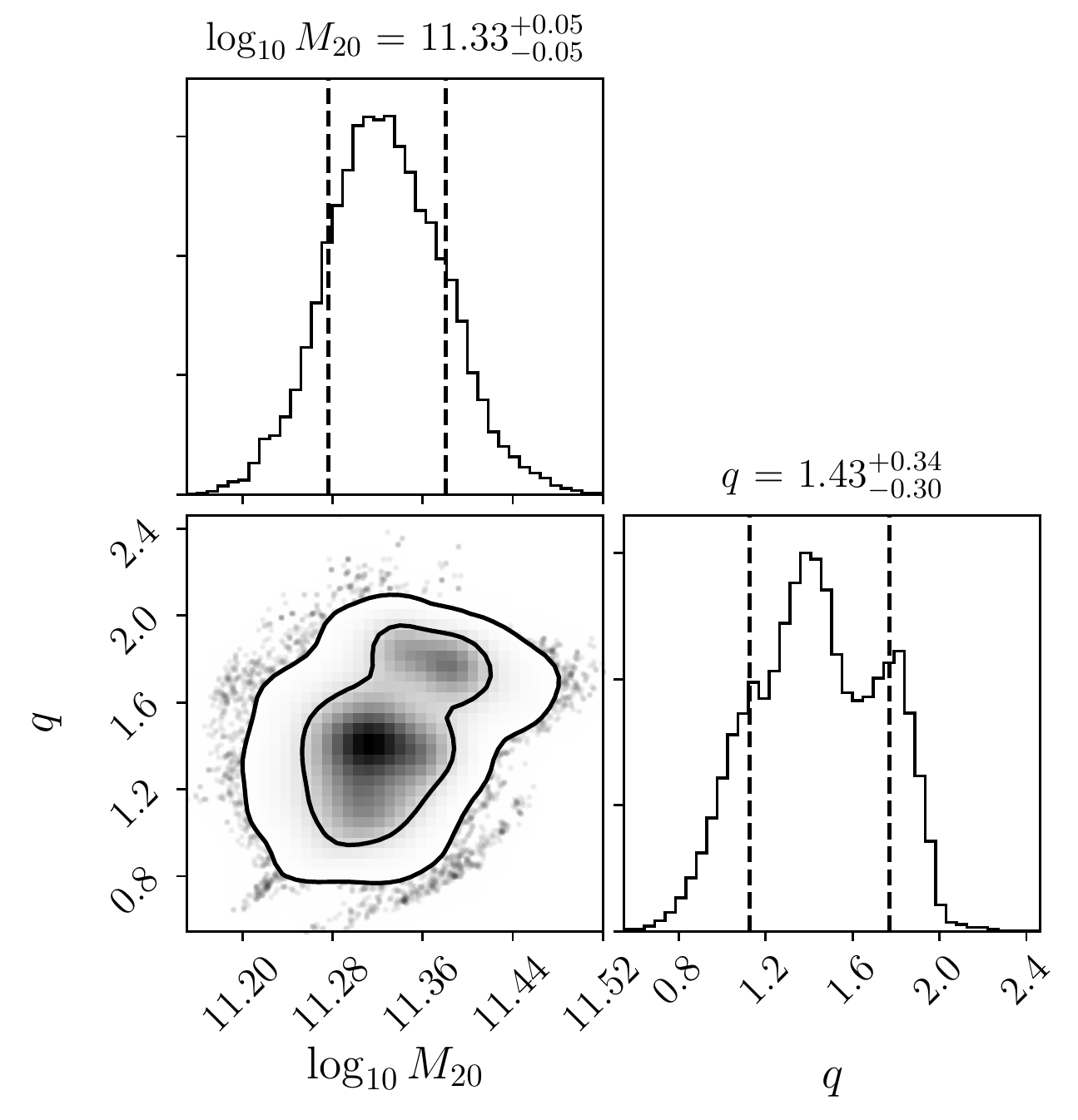}
\caption{Posterior distributions of the two parameters describing the Galaxy's dark
                 halo: the total mass of the Galaxy within 20 kpc, $\Mtw$, and the 
         axis ratio of the dark halo, $q$. The number of bins is increased with respect to Fig.~\ref{fig:mcmc_full}.
         Black solid curves and vertical dashed lines are as in Fig.~\ref{fig:mcmc_full}.
        }
\label{fig:corner_pot}
\end{figure}

With our modelling technique we are able to constrain the characteristic parameters
of the Galaxy's inner dark matter halo that fit the observed dynamics of GCs.
The main novelty of our work is the fact that the new dataset released by the
\cite{Gaia+18b} gives us strong inference on both the total mass and shape
of the dark matter halo. In particular, in Figure \ref{fig:corner_pot} we show 
the posterior distribution of the Galaxy's mass within 20 kpc and the halo's
axis ratio $q$ as sampled by our MCMC analysis. 
We find that $\logten\Mtw/\Msun=11.33\pm 0.05$ and
$q=1.43^{+0.34}_{-0.30}$ and that they are strongly correlated with more massive
halos being more prolate. While such a degeneracy is  expected
\citep[e.g.][]{Bowden+16}, here we nonetheless find that spherical and light
halos are disfavoured by our analysis, while halos more oblate than $q\leq 0.8$ 
are ruled out with $99\%$ probability by our experiment, as are strongly prolate
halos with $q\geq 1.9$.

The posterior distribution that we sample is  not unimodal, but it has
two distinct peaks;  the most likely is at $\logten\Mtw\simeq 11.33$, $q\simeq 1.37$
and the other at $\logten\Mtw\simeq 11.38$, $q\simeq 1.75$. Although it  is located
in a region where the likelihood is considerably lower, the second peak
cannot be neglected and deserves further investigation.

\begin{figure}
\includegraphics[width=0.49\textwidth]{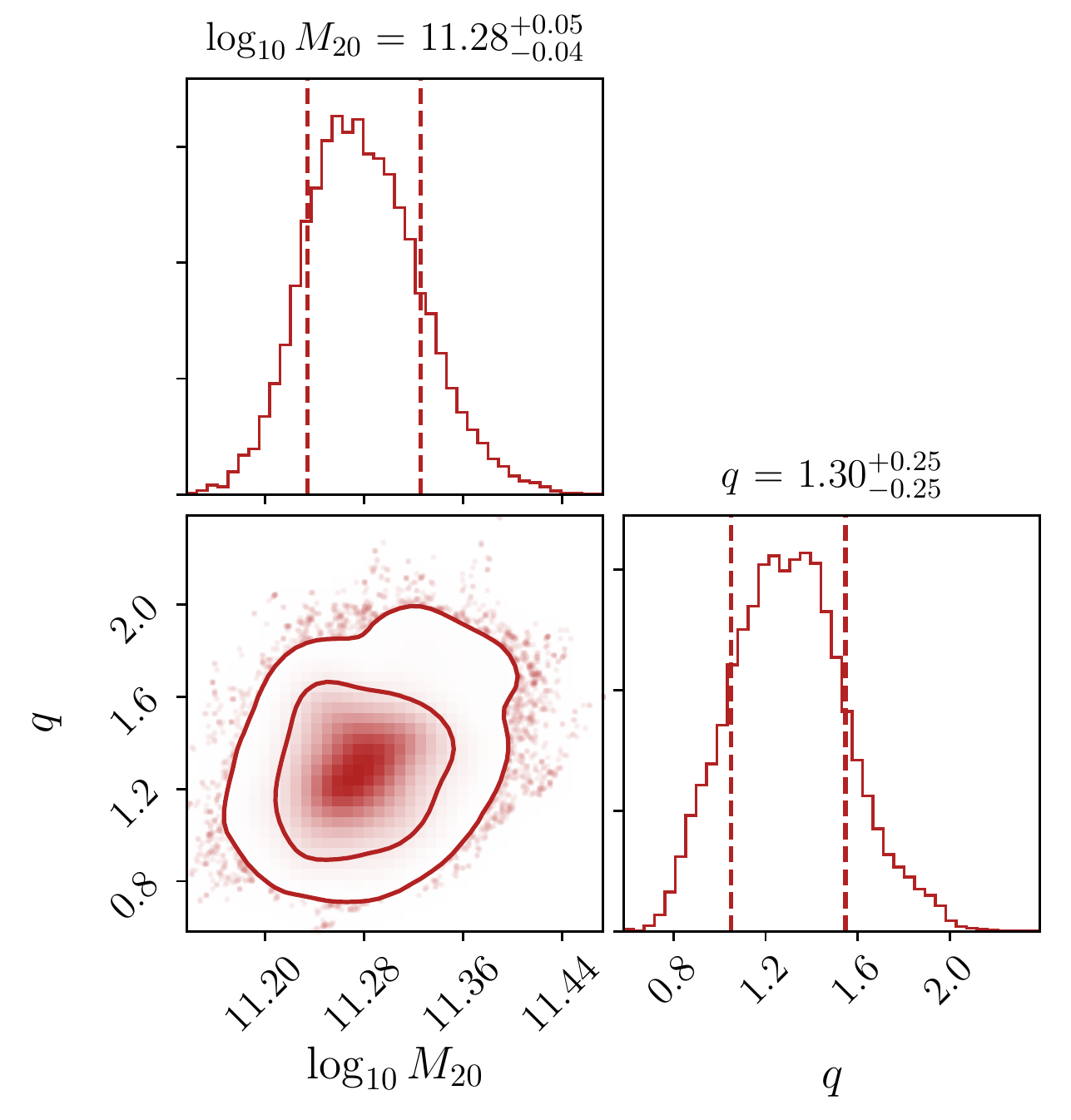}
\caption{Same as Fig.~\ref{fig:corner_pot}, but obtained excluding the GCs
                 associated with the Sagittarius dwarf galaxy.
        }
\label{fig:corner_pot_noSag}
\end{figure}

As discussed in the introduction, an important number of GCs in our sample are 
likely associated with the Sagittarius dwarf galaxy and have thus been accreted on 
similar orbits. In this case, our assumption that all the GCs are a random sample of
our DF breaks down, and this could lead to biases in the results of the modelling.
We therefore remove the clusters mentioned in Sec.~\ref{sec:sgr}, and repeat the analysis 
outlined in Sect.~\ref{sec:mcmc}. As a result we obtain 
the posterior distribution for the halo parameters shown in Figure
\ref{fig:corner_pot_noSag}. Now the second lower peak has completely disappeared and the 
distribution has a well-constrained peak at $\logten\Mtw/\Msun=11.28\pm 0.04$ and
$q=1.22 \pm 0.23$. From this we conclude that the second peak was most probably driven
by the few clusters on very polar orbits dynamically associated with the Sagittarius dwarf
galaxy.

Within 20 kpc, we find the total mass of the Galaxy to be $1.91^{+0.18}_{-0.17}
\times 10^{11}\Msun$, which implies a dark matter mass of $1.37^{+0.18}_{-0.17}\times
10^{11}\Msun$.
If we extrapolate this to the virial radius\footnote{
We adopt the \cite{BryanNorman98} definition of the virial radius, with a virial
overdensity at redshift zero of $\Delta\simeq 102.5$.
}, assuming a concentration-mass relation as calibrated from numerical simulations
\citep{DuttonMaccio14}, we obtain the following final estimate of the Galaxy's dark
matter halo mass,
\begin{equation}\label{eq:virialmass}
\Mvir = 1.3 \pm 0.3 \times 10^{12}\Msun
\end{equation}
and thus a virial radius of $r_{\rm vir}=287^{+22}_{-25}$ kpc.

Many different studies have already estimated the mass of the dark matter halo of
the Milky Way using very different techniques and tracers, which
typically are sensitive to very different physical scales. Thus, a proper
comparison taking into account all the possible biases given by the different
assumptions or data used is needed, but it goes beyond the scope of the present study.
By naively comparing our results with the numerical estimates compiled by
\citet[][and references therein]{BlandHawthornGerhard16} and \citet[][and
references therein]{EadieHarris16} we conclude that our estimate of $\Mtw$ and
of the extrapolated $\Mvir$ lies somewhere between the very light
\citep[$\Mvir<10^{12}\Msun$, e.g.][]{Battaglia+05,Gibbons+14} and of the very
heavy values \citep[$\Mvir>2 \times 10^{12}\Msun$][]{LiWhite08,Watkins+10}.
In general, we find good agreement with other works that use satellite kinematics
\citep[provided that distant enough tracers are included, e.g.][]{BoylanKolchin+13,
Gonzalez+13,EadieHarris16,Patel+18,Sohn+18}, estimates of the escape velocity
\citep[e.g.][]{Smith+07,Piffl+14a}, 
the velocity distribution of either fast moving stars \citep[e.g.][]
{Gnedin+10}, or the dynamics of nearby stars  \citep[e.g.]
[]{McMillan11,Piffl+14}.

Similarly, there has also been  some work on the shape of the total gravitational
potential. While a proper comparison taking into account the systematics induced
by tracers and methods is not straightforward, here we note that our estimate roughly
agrees with other  studies of the Sagittarius stream \citep{Helmi04}, the
flaring of the HI disc \citep{BanerjeeJog11}, or the kinematics of halo stars
\citep{Bowden+16}, while being apparently inconsistent with studies modelling the GD-1
stream \citep[][which, however, probes a region much closer to the Galactic centre]
{Koposov+10,Bowden+15} and studies of the tilt of the local velocity ellipsoid \citep{Smith+09,
Posti+17}.

It may well be the case that the shape of the halo of our Galaxy is considerably
more complicated than  we have assumed here; for instance, i) it can be axisymmetric,
but with a variable flattening as a function of radius \citep[e.g.][]{VeraCiro+13}, 
being oblate close to the disc and more spherical/prolate at greater distances 
\citep[e.g. as probed by the HI disc flare,][]
{BanerjeeJog11}; ii) it can be triaxial \citep[e.g. as probed by the modelling of the 
Sagittarius stream][]{LawMajewski10}; or iii) it can be triaxial, but with variable axis
ratios/orientations as a function of radius \citep[e.g. as expected from numerical
simulations,][and as probed by the Sagittarius stream, \citeauthor{VeraCiro+13} 
\citeyear{VeraCiro+13}]{BailinSteinmetz05,VeraCiro+11}.
Different results are thus found both because of different model assumptions and 
because of the various observables used: since different tracers are probably
sensitive to distinct portions of the halo, the methodologies used to infer
the halo shape may be subject to different kinds of limitations. This implies 
that to make further progress in our understanding of the shape of the dark halo
of the Milky Way it is imperative to combine the signal coming from tracers of
different nature.

\section{Conclusions}\label{sec:concl}

We have used the kinematics of Milky Way GCs to  simultaneously constrain the
phase-space distribution of the GC system, the mass of the inner Galaxy, and
the shape of its inner dark matter halo. This was possible thanks to the novel
measurements by \Gaia~and HST of the proper motions of 91 GCs with  unprecedented 
accuracy. 

Our method is based on the assumption that 
the GC system can be described by axisymmetric equilibrium models with two
distribution functions, which  are analytic functions of the action integrals, 
representing the phase-space distribution of disc and halo clusters. 
Our models include an axisymmetric NFW dark matter halo, whose mass and density axis
ratio, have been determined by fitting the kinematics of 143 GCs  (91 with accurate
6D data and 52 with only radial velocities), using a Bayesian approach, imposing
uninformative priors on the eight model characteristic parameters.

We summarise here our main results:
\begin{itemize}
\item[(i)] we find all the parameters of the halo component to be  well constrained.
           The density distribution is well described by a single power law
           $\rho\propto r^{-3.3}$; the velocity distribution is mildly radially
           biased, with a rather constant anisotropy $\beta\simeq 0.2\pm 0.07$,
           and consistent with either a small counter-rotation ($V_{\rm rot}\simeq -15\,
           \rm km\,s^{-1}$ at 20 kpc) or no net rotation;
\item[(ii)] we find all the parameters of the disc component to be  well constrained.
            Their phase-space distribution closely resembles that of the thick disc
            of the Galaxy \citep{Piffl+14};
\item[(iii)] metal-rich clusters ([Fe/H]$>-0.8$) very likely belong to the disc
             component, while roughly $67\%$ of the metal-poor clusters
             ([Fe/H]$\lesssim-0.8$) are more likely part of the halo component.
             All clusters with low binding energies belong to the halo component 
             and have both prograde and retrograde orbits. The more bound halo clusters 
             are typically found on counter-rotating or non-rotating orbits. Most metal-rich
             clusters ([Fe/H]$\gtrsim-0.8$) have high a probability of   being associated with the
             disc component. Our findings are broadly consistent with the picture
             proposed by \cite{Zinn85};
\item[(iv)] we measure the mass of the dark matter halo of the Galaxy within
            20~kpc to be $\logten M_{20,\rm DM}/\Msun=11.14\pm 0.05$. This very
            accurate measurement implies a total virial mass for the Milky Way of
            $\Mvir = 1.3 \pm 0.3 \times 10^{12}\Msun$ after assuming a 
            concentration-virial mass relation; 
\item[(iv)] we measure a flattening $q=1.22 \pm 0.23$, and find oblate halos with
                        $q < 0.7$ to be ruled out by our analysis (with $99\%$ probability)  and 
            spherical models to be disfavoured in comparison to prolate halos. 
\end{itemize}

During the completion of this work, \cite{Watkins+18} presented an estimate of the
Milky Way's dark halo mass by applying a simple `tracer mass estimator' on a similar
dataset. Their values for  the mass within 21.1 kpc (and within the virial radius) are
closely compatible with ours, as is their measurement of the anisotropy (although they 
favour a slightly more radially biased ellipsoid).

Our work demonstrates that with a giant leap in data quality such as that provided
by the \Gaia~DR2, constraints on fundamental physical quantities become significantly
more precise, and several of these parameters can be determined at once if  
models with enough sophistication are employed. Of course, with this work we have simply touched
the tip of the iceberg of the signal present in the \Gaia~data regarding the mass
and shape of the Galactic potential.
The next challenge will be to simultaneously fit the dynamics of different mass tracers
such as  satellites, streams, field stars in the halo, and hypervelocity stars. 
This work may be seen as a necessary first step in this direction,
and merely confirms that we may have just entered the era of `Precision Galactic
Astronomy'.

\begin{acknowledgements}
We thank the referee, Matthias Steinmetz, for a constructive report and Simon White
for useful discussions.
We acknowledge  financial  support from a VICI grant from the Netherlands
Organisation for Scientific Research (NWO).
This work  has  made  use  of  data  from  the  European  Space  Agency  (ESA)  
mission Gaia (\url{http://www.cosmos.esa.int/gaia}), processed by the Gaia Data
Processing and Analysis Consortium (DPAC,
\url{http://www.cosmos.esa.int/web/gaia/dpac/consortium}).
Funding for the DPAC has been provided by national institutions, in particular
the institutions participating in the Gaia Multilateral Agreement.
\end{acknowledgements}

\end{document}